\documentclass[epsf,twocolumn,showpacs,preprintnumbers]{revtex4}
\usepackage{graphics}
\usepackage{graphicx}
\usepackage{dcolumn} 
\usepackage{bm}
\usepackage{epsfig}
\begin{document}
\title{Crystal Symmetry, Electron-Phonon Coupling,
    and Superconducting Tendencies in Li$_2$Pd$_3$B and Li$_2$Pt$_3$B} 
\author{K.-W. Lee and W. E. Pickett} 
\affiliation{Department of Physics, University of California, Davis, 
  CA 95616, USA}
\date{\today}
\begin{abstract}
After theoretical determination of the internal structural coordinates in
Li$_2$Pd$_3$B, we calculate and analyze its electronic structure and obtain
the frequencies of the two $A_g$ phonons (40.6 meV for nearly pure Li mode,
13.0 meV for the strongly mixed Pd-Li mode).
Pd can be ascribed a $4d^{10}$ configuration, but strong $4d$ character
remains up to the Fermi level. 
Small regions of flat bands occur at $-70$ meV at both the $\Gamma$ and X points.
Comparison of the Fermi level density of states to the linear specific heat
coefficient gives a dynamic mass enhancement of $\lambda$ = 0.75. 
Rough Fermi surface averages of the deformation potentials ${\cal D}$ of individual
Pd and Li displacements are obtained.  While $\langle{\cal D}_{\tt Li}\rangle$ is small,
$\langle{\cal D}_{\tt Pd}\rangle \approx$ 1.15 eV/\AA~is sizable, and a plausible case 
exists for its superconductivity at 8 K being driven primarily by coupling to Pd
vibrations.  The larger $d$ bandwidth in Li$_2$Pt$_3$B leads to important differences 
in the bands near the Fermi surface.  The effect 
on the band structure of
spin-orbit coupling plus lack of inversion is striking, and is much
larger in the Pt compound.
\end{abstract}

\maketitle

\section{introduction} 
 The occurrence of superconductivity in compounds without inversion
symmetry is an issue that has surfaced recently.\cite{bauer}
Although there seems to be very few compounds without a center
of inversion which display robust
superconductivity, a look at basic BCS theory reveals no severe obstruction
to pairing, as long as there is no ferromagnetism, because 
time-reversal invariance (i.e., lack of magnetism) is sufficient to
guarantee inversion symmetry in the Fermi surface 
($\varepsilon_{-k\sigma}$=$\varepsilon_{k\sigma}$), therefore allowing zero
momentum ($q=0$) pairing.  Ferromagnetism splits the spin ($\sigma$)
degeneracy, restricting $q=0$ pairing to triplet pairs.
If inversion symmetry is lost, however,
spin-orbit coupling (SOC) removes the 
spin degeneracy\cite{samokhin,sergienko} so that
$q=0$ pairing may be very sensitive to magnetism (including applied
magnetic fields).\cite{samokhin2,frigeri0}

The lack of inversion symmetry has been suggested as a factor in
the absence of superconductivity in MnSi\cite{pfleiderer,frigeri}; there are however
other peculiar aspects of the space group 
(and resulting band structure)\cite{jeong}
that hamper elucidation of the effects of lack of inversion.
Very recently a few examples of superconductivity in crystalline materials
without inversion symmetry have been reported.
CePt$_3$Si, with a crystal structure that is not close to any
structure with inversion, has been found to be superconducting
at T$_c$ = 0.75 K.\cite{bauer}  The properties of this system are complicated
due to its heavy fermion nature.  SOC effects might be expected to cause 
complications\cite{sergienko} (due to the cerium) 
and the fact that superconductivity arises
within the antiferromagnetic phase (T$_N$ = 2.2 K) with enforced spin
degeneracy may be relevant.
Pressure induced superconductivity at P$_c$ = $2.6$ GPa 
and very low T$_c$ = 0.14 K
has been observed in UIr.\cite{akazawa}  Electrons in UIr will experience
strong SOC, however the crystal
structure is only a small distortion away from one with inversion, so
the effects of lack of inversion can only be determined 
by direct calculation.

Recently Li$_2$Pd$_3$B with a cubic but peculiar crystal structure
without inversion symmetry has been discovered to be a T$_c$ = 8 K
superconductor by Togano and coworkers.\cite{togano1,togano2}
It has an upper critical field H$_{c2}$ = 4 T, and 
the Ginzburg-Landau parameter $\kappa$ = 21, marking it as a strongly type-II
superconductor.\cite{togano2}
Although there has been a suggestion\cite{sardar} the superconductivity is 
dominated by strong electronic correlations related to three kinds of 
the Pd-Pd bond length,\cite{jung} 
experimental data is readily interpreted without any need for invoking 
correlation effects.\cite{togano3,nishiyama}  Moreover, the density of 
states presented by Chandra {\it et al.} indicates that the Pd $4d$ bands
are essentially completely filled,\cite{chandra} leaving no reasonable
expectation of correlation effects on the Pd site.
The entire alloy system Li$_2$B(Pd$_{1-x}$Pt$_x$)$_3$ has been
studied, and it was found\cite{togano4} that
T$_c$ decreases almost linearly 
from 8 K (Pd end) to 2.8 K (Pt end).  Since the volume is unchanged 
(the lattice constant of Li$_2$Pt$_3$B is 0.03\% larger), the difference
is due to (1) the slightly different chemistry of Pt, (2) the mass
difference, or (3) the effect of stronger SOC combined with the lack of
inversion symmetry of this lattice.  Interpreted as an isotope shift
$\alpha$= $-$d(log T$_c$)/d(log M) leads to $\alpha$ greater than unity,
although changes in both T$_c$ and the mass M are too large for the
differential definition of $\alpha$ to be realistic.

The Li$_2$Pd$_3$B structure has been described as a three-dimensionally distorted
antiperovskite,\cite{jung} characterized by strongly 
distorted corner-sharing BPd$_6$
octahedra.  Thus on the local structural level Li$_2$Pd$_3$B appears to have
similarity to MgCNi$_3$,\cite{he} which is also an 8 K superconductor and
has a similar valence balance (one more valence electron per formula unit).  The
lack of inversion symmetry is however only one aspect of the strong
difference between the structures of Li$_2$Pd$_3$B and MgCNi$_3$.  
The structure
and space group are discussed in some detail in Sec. II.
There are several Pd and Pt based (anti)perovskite compounds\cite{cava}
that may be more strongly related to Li$_2$Pd$_3$B, but they have not
been studied nearly so thoroughly.

In this paper, we investigate in detail the electronic structure of 
Li$_2$Pd$_3$B and its relation to the local bonding and to the global
symmetry of the crystal structure.  We also begin some investigation 
into the lattice dynamics and electron-phonon coupling by studying the
symmetric vibration of the Li and Pd atoms.  We obtain rough estimates 
of the contribution of Li and Pd motions to electron-phonon coupling
strength, and obtain a plausible case that Pd motion is the primary
driver of superconductivity.

\section{Structure and Calculation Method}
Our calculations were based on the experimentally reported structure\cite{jung}
(cubic $P$4$_3$32, No. 212), containing four formula units per primitive
cubic cell, using the lattice constant 6.7436~\AA~ obtained by 
Togano {\it et al.}\cite{togano1}
The peculiarity of the structure is
already evident at a basic level: the simple cube cell contains {\it four}
BPd$_6$ octahedra.  This is not possible by simply enlarging the cubic
perovskite cell, as its cubic supercells contain 8, 27, ... $p^3$ octahedra
for a $p\times p\times p$ supercell.  Thus the {\it topology} of the
BPd$_6$ octahedra network is distinct from a perovskite such as MgCNi$_3$.
This space group consists of a threefold axis without any associated
nonprimitive translation; all other rotations are paired with a
($\frac{1}{2},\frac{1}{2}$,0), ($\frac{1}{4},\frac{1}{4},\frac{1}{4}$),
or ($\frac{3}{4},\frac{3}{4},\frac{1}{4}$) type translation.
Lack of inversion leaves 24 operations in the point group.

\begin{table}[bt]
\caption{Significant interatomic distances (in \AA) after relaxation.
 Compared with the experimentally reported value,\cite{jung} ($\uparrow$) and
 ($\downarrow$) indicate increase and decrease, respectively.  Note
that all B-Pd nearest neighbor distances are identical.}
\begin{center}
\begin{tabular}{ll}\hline\hline
  Pd$-$Pd~~ & 4$\times$2.779($\downarrow$),~2$\times$2.989($\uparrow$),~
          2$\times$3.540($\uparrow$) \\
  Pd$-$B~~  & 2$\times$2.135($\uparrow$),~1$\times$4.113($\uparrow$) \\
  Pd$-$Li~~ & 2$\times$2.693($\downarrow$),~2$\times$2.838($\uparrow$),~
          2$\times$3.864($\uparrow$) \\
  Li$-$Li~~ & 3$\times$2.584($\downarrow$) \\\hline\hline
\end{tabular}
\end{center}
\label{table1}
\end{table}

Pd atoms lie at $12d$ sites ($\frac{1}{8},x_1,\frac{1}{4}-x_1$) with very low
(twofold rotational) symmetry, with reported $x_1=0.30417$.
Li atoms reside at $8c$ sites ($x_2,x_2,x_2$) with threefold symmetry, 
with reported $x_2=0.3072$.\cite{jung} 
B atoms within Pd octahedra lie on $4b$ sites 
($\frac{5}{8},\frac{5}{8},\frac{5}{8}$) 
and have ``32'' (threefold+twofold) symmetry.
While the B atom lies at the center of mass (CM) of the Pd octahedron and
the B-Pd distances are all equal (2.135~\AA), the octahedron is strongly
distorted.  The Pd-B-Pd bond angles are 162$^{\circ}$ that would be
180$^{\circ}$ in a cubic perovskite, and the 90$^{\circ}$ angles
become 81$^{\circ}$, 89$^{\circ}$, and 112$^{\circ}$.
The distortion keeps one pair of opposing faces of the octahedron
perpendicular to the local threefold $<111>$ direction.

The B sublattice consists of symmetry-determined sites, which for the
four in the primitive cube lie at ($1,1,1$), ($-3,3,-1$), ($-1,-3,3$), and
($3,-1,-3$) in units of $\frac{a}{8}$ and with respect to CM of the four.  
Each B lies on a threefold axis, with B-CM-B angles of 97.6$^{\circ}$ 
and 118.3$^{\circ}$, so they are not tetrahedrally placed with respect
to their CM.
This arrangement results in each B atom lying at a connecting vertex of
{\it three} equilateral triangles of B atoms with side length $4.1356$~\AA. 
This local connectivity makes the B sublattice resemble a three dimensional
generalization of the Kagome lattice, that is, a 3D network of 
interconnected triangles.

Due to the low xray scattering cross section for Li atoms, the Li
position may not have been accurately determined. 
The two internal parameters mentioned above (Li and Pd)
were relaxed within the local density 
approximation, using the methods discussed below. After relaxation, 
we obtain for Pd $x_1=0.3057$ (different by $+0.014$~\AA) and for Li 
$x_2=0.3018$ (different by $-0.063$~\AA).  This difference for Li may be
large enough to have significance for the electronic structure.
The resulting interatomic distances are given in Table \ref{table1}.

 The calculations were done with the full-potential nonorthogonal
local-orbital minimum-basis method (FPLO).\cite{fplo1}
The fully relativistic scheme,\cite{fplo2} implicitly equivalent to
the spin-orbit coupling, implemented in FPLO was also used.
Valence bands included Pd $4s4p5s5p4d$, Li $1s2s2p3d$, and B $2s2p3d$.
The Brillouin zone was sampled with 200 ($16\times16\times16$) 
irreducible $k$ points.

\section{Results and Interpretation}
\subsection{Electronic structure}
\begin{figure}[tbp]
\rotatebox{-90}{\resizebox{7cm}{8cm}{\includegraphics{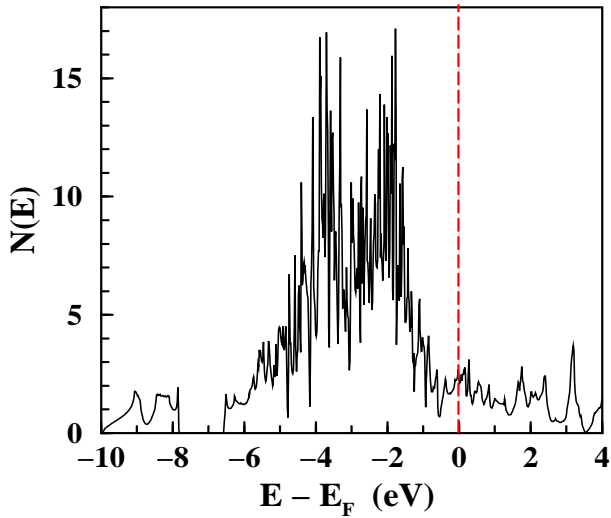}}}
\caption{Total DOS for both spins per a formula unit of Li$_2$Pd$_3$B. 
 Every occupied band, except B $2s$ band lying on $-10$ eV to $-8$ eV, 
 is strongly mixed above $-6.7$ eV.
 But, states from $-6$ eV to $-1$ eV are mostly Pd $4d$ states.
 The vertical dashed line indicates the Fermi energy.}
\label{dos}
\end{figure}

\begin{figure}[tbp]
\rotatebox{-90}{\resizebox{7cm}{8cm}{\includegraphics{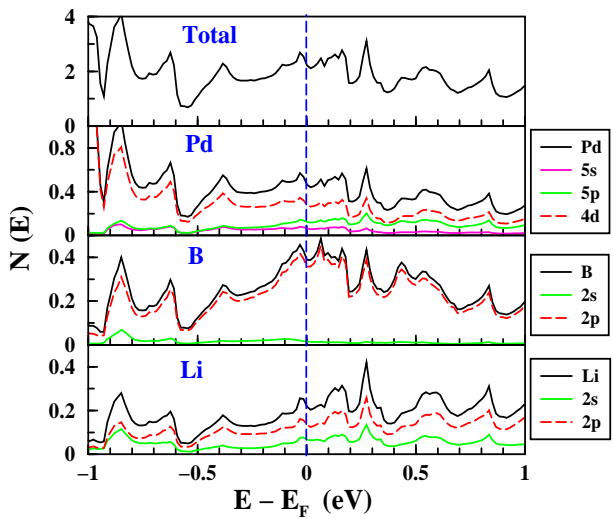}}}
\caption{(Color online)Atomic and orbital projected DOS per atom 
 near the Fermi energy. (The total DOS is given for per eV per 
 formula unit.) 
 $N(0)$ is decomposed into Pd 60\%, Li 20\%, and B 20\%, but 
 in particular Pd $4d$ 50\%.
 Note different scale of $N(E)$ for each plot.
 The vertical dashed line indicates the Fermi energy.}
\label{pdos}
\end{figure}

The density of states (DOS) of the full valence band 
is given in Fig. \ref{dos}
and agrees with the result of Chandra et al.\cite{chandra}
B $2s$ states are separate in the $-10$ eV to $-8$ eV range.
The other occupied bands are a mixture of B $2p$, Li $2s$ and $2p$, 
and Pd $4d$ states.
The main complex of Pd $4d$ states extends from $-6$ to $-1$ eV,
indicating a $d^{10}$ description is most appropriate,
but the $4d$ character tails up to 4 eV owing to hybridization with 
B $2p$ states.  The low internal symmetry and twelve Pd atoms per cell
lead to 60 Pd $4d$ bands in the 5 eV range, giving rise to the
``hairy'' DOS in Fig. \ref{dos}.
As a result of the strong hybridization, the B $2p$ states
are themselves repelled, so that the B $2p$ states are divided
into one region (containing about 20\% of the states) from $-6.7$ eV to 
$-4$ eV and another region above $-1$ eV.
The Li $2p$ character is also separated into one region (containing
10\% of the states) from $-6.7$ eV to about $-1.3$ eV and another
region above $-1$ eV.
Separations due to this repulsion make electron-depleted 
deep valleys, especially around $-4$ eV and $-1$ eV.
The Li $2s$ states spread across a wide range, but they are only
about 20\% occupied
(i.e. 1/5 and not 1/2 of the isolated Li atom).
Thus Li may be somewhat cationic in this compound, a conclusion reached
by Chandra {\it et al.} on geometric grounds.\cite{chandra}

The band mixture remains complex near the Fermi energy (E$_F$), as shown by 
the atomic and orbital projected DOS in Fig. \ref{pdos} in a small
region near E$_F$ (taken as the zero of energy).  The total and the Pd
contribution are relatively constant within a few tenths of an eV of E$_F$,
whereas the B $2p$ contribution has a maximum near E$_F$.
The DOS at the Fermi level $N(0)$ is 2.24 states/eV per formula unit
(both spins).\cite{Nf}
It is composed of 60\% Pd, 20\% Li, and 20\% B characters (note that the 
contribution {\it per atom} is almost the same from B as from Pd).
In particular, the Pd $4d$ states show the primary contribution, about 50\% 
of $N(0)$, even though the ``full $d$ bands" are best pictured 
in terms of a $4d^{10}$ configuration. 
The linear specific heat coefficient $\gamma$ = 9.0 mJ/mol-K$^2$\cite{togano5}
corresponds to 
a quasiparticle density of states $N^*(0) \equiv (1+\lambda)N(0)$ = 
3.82 states/eV-formula unit.  This leads to a dynamical mass enhancement 
$\lambda$ = 0.75.

\begin{figure}[tbp]
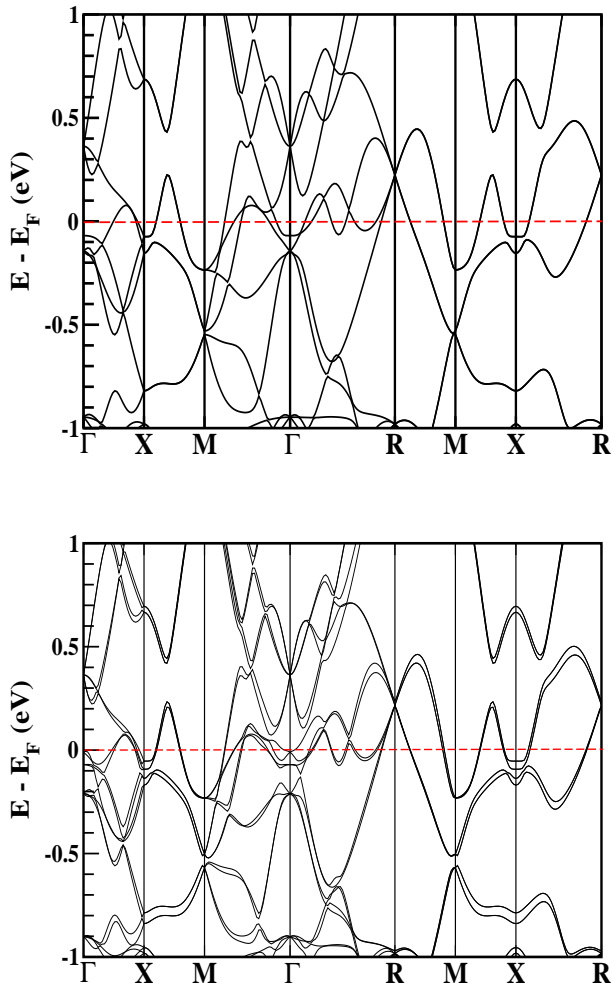

{\resizebox{8cm}{6cm}{\includegraphics{Fig3a.eps}}}
\vskip 10mm
{\resizebox{8cm}{6cm}{\includegraphics{Fig3b.eps}}}
\caption{Blowup of the band structure within 1 eV of the Fermi energy.
 Note that the flat bands lie at $-70$ meV at the $\Gamma$ and X points
in the bands without spin-orbit coupling included (top panel).
The bottom panel includes spin-orbit coupling (see text for discussion).
R denotes the zone boundary point along a $<111>$ direction.
 The horizontal dashed lines indicate the Fermi energy.}
\label{band}
\end{figure}

\begin{figure}[tbp]
{\resizebox{8cm}{6cm}{\includegraphics{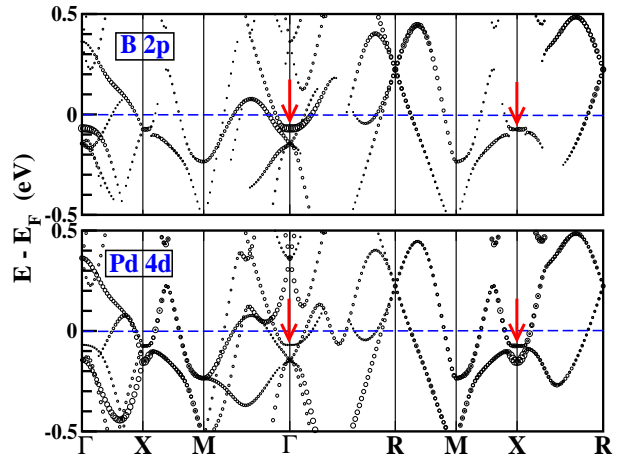}}}
\caption{(Color online)``Fatband" representations of the B $2p$ 
 and Pd $4d$ states within $0.5$ eV of the Fermi energy.
 (Li shows very little character in this range.)
 The size of symbols is proportional to character of B $2p$ (Pd $4d$)
 states.  
 The arrows denote the flat bands lying at $-70$ meV at the $\Gamma$
 and X points.
 The horizontal dashed lines indicate the Fermi energy.}
\label{fat}
\end{figure}

The complexity of the band structure is displayed clearly in the expanded
band structure near E$_F$ in Fig. \ref{band}. 
Four bands cross E$_F$, but the most interesting feature
comes from as small region of flat bands lying just $-70$ meV below E$_F$ 
at both the $\Gamma$ and X points. The flat bands are sensitive 
to the position of the atoms, indicative of electron-phonon coupling (see below). 
In contrast to the result of the previous report\cite{chandra} using
the experimental values of the Li position, the flat bands are
located at the identical energy of $-70$ meV after relaxation,
because the flat bands at the
$\Gamma$ and X points have different deformation potentials (Sec. III.D).
The difference in the deformation potential comes from difference in
character of the flat bands, as shown in the fatband representations
of Fig. \ref{fat}.
Both flat bands are B $2p$ and Pd $4d$ mixture.
However, the band at the $\Gamma$ point is mostly B $2p$ character,
while at the X point Pd $4d$ and B $2p$ characters have similar magnitude. 

In addition, the nonsymmorphic space group without inversion leads to
unusual behavior in the band structure, 
as found in MnSi with the B20 structure\cite{jeong} which also has a
cubic Bravais lattice (but a quite distinctive one).
The most clearly evident result of the nonsymmorphic nature of 
this $P$4$_3$32 space 
group is that it leads to all bands at the zone 
corner R point being fourfold degenerate,
see for example the complex at 0.2 eV in the top panel of Fig. \ref{band}.
Secondly, many bands at symmetry points have the unusual feature 
of nonzero velocity
due to the space group; see for example the aforementioned bands at the R
point, some threefold representations at the $\Gamma$ point which have 
two nonzero velocities (at 0.3 eV in the top panel of Fig. \ref{band}), 
and also some bands at the X and M points.  
This type of ``band sticking'' due to non-symmorphic operations 
has been discussed at some length for the case of MnSi.\cite{jeong} 
Note that the threefold bands at the $\Gamma$ point consist of one band 
with zero velocity and two other bands with nonzero velocities 
having identical magnitude but opposite sign.

\subsection{Fermi Surfaces}
\begin{figure}[tbp]
{\resizebox{8cm}{6cm}{\includegraphics{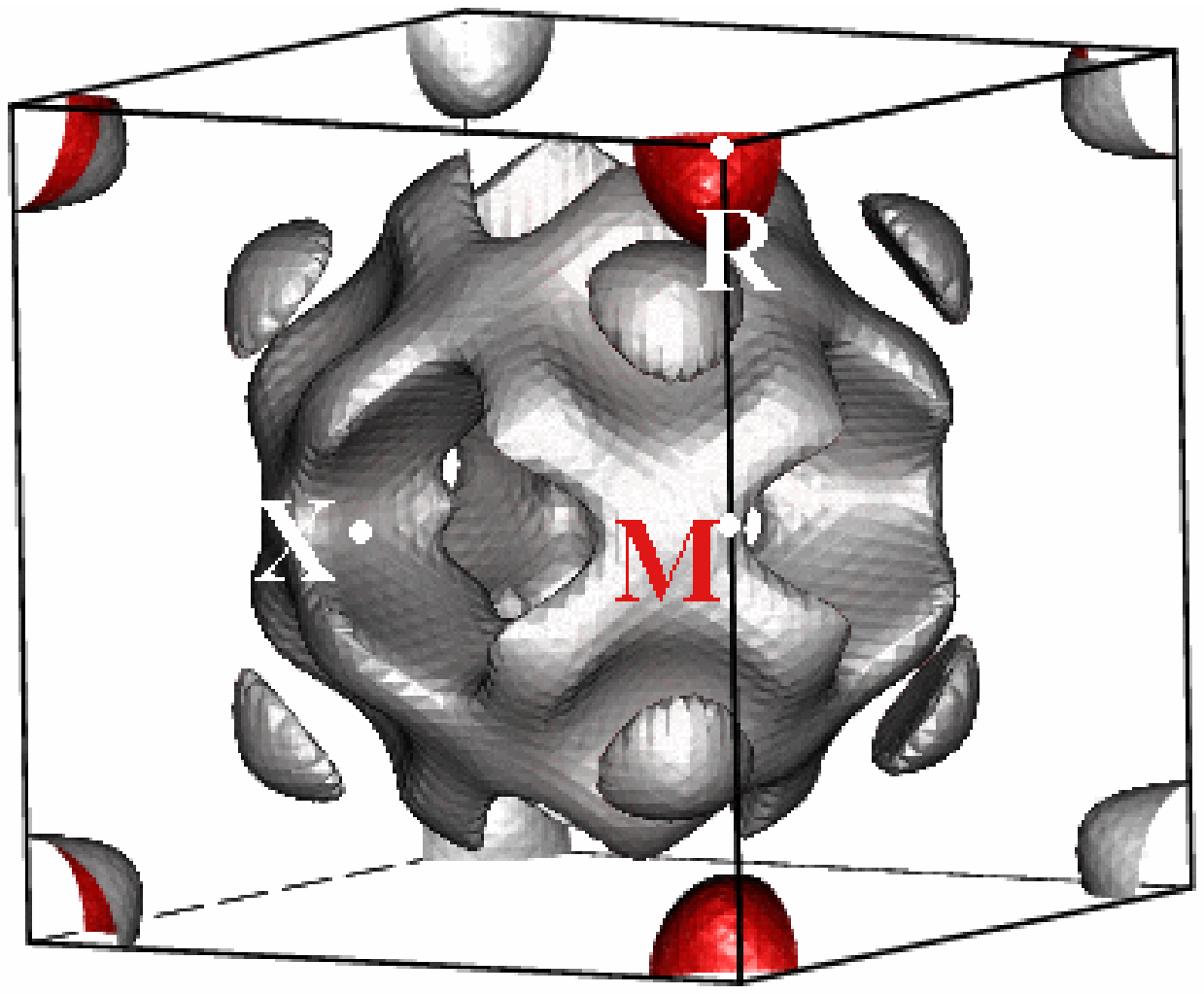}}}
\vskip 10mm
{\resizebox{8cm}{6cm}{\includegraphics{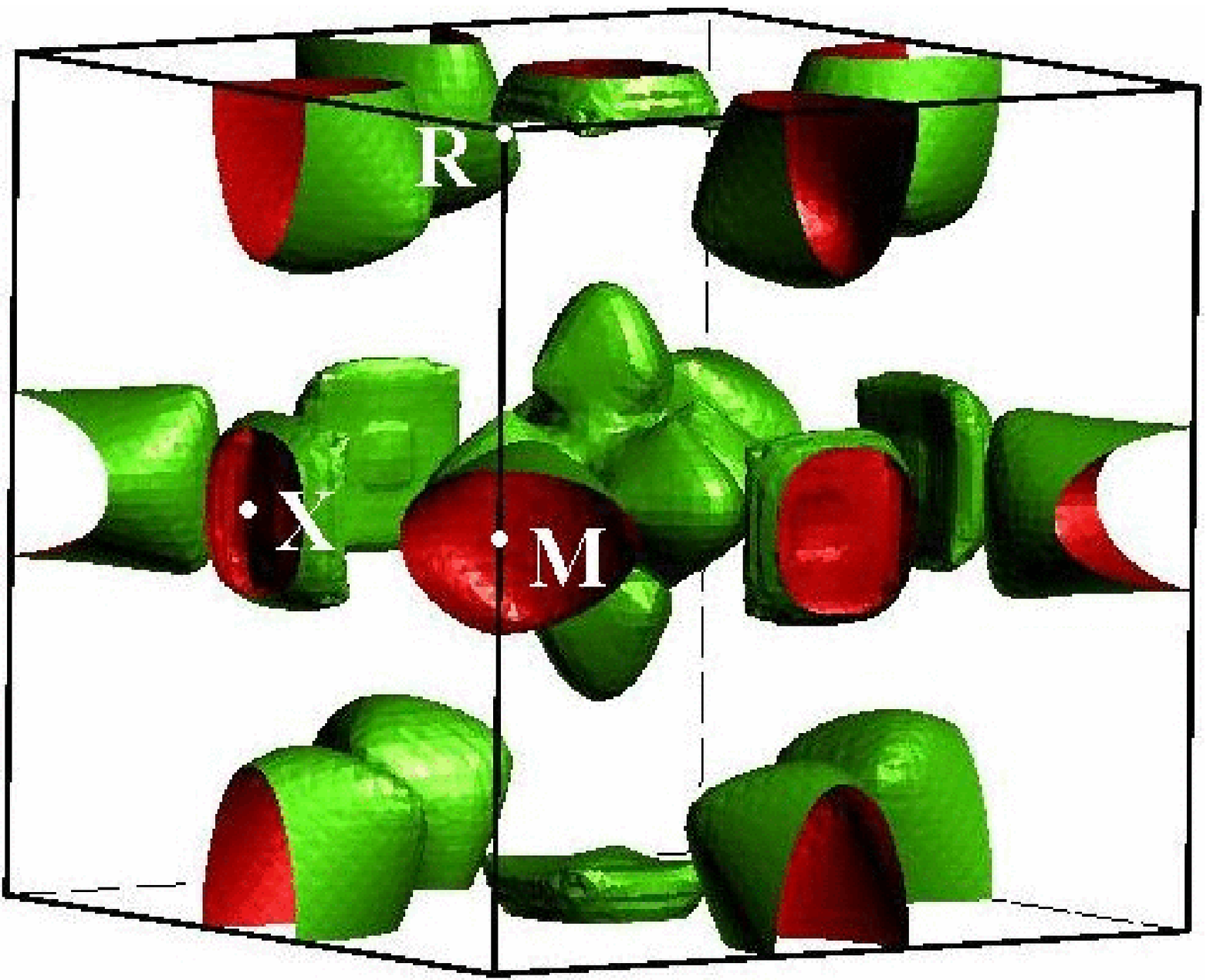}}}
\caption{(Color online) Fermi surfaces of Li$_2$Pd$_3$B,  
 with positions of the symmetry points labeled.  Several ellipsoid
 pieces are not shown here, but are noted in the text.  The Fermi surfaces
shown are from the second and third of the four bands that cross the 
Fermi energy and are described in the text. }
\label{FS}
\end{figure}

We show the more complex pieces of the Fermi surface in Fig. \ref{FS} that 
correspond to the bands shown in the top panel of Fig. 3.  In addition to
those shown, there are ellipsoid-like pockets at the X (electron), R (hole), 
and M (electron) points.  
The dimensions of these simple pockets can be seen from Fermi level 
crossings in Fig. 3.

The top panel of Fig. \ref{FS} shows a $\Gamma$-centered surface that is
topologically that of a hollow ball with holes along the cubic axes
(``wiffle ball'').  The
topology of this surface, which encloses holes, is quite involved, with
areas of both positive and negative curvature.  In addition there are 
lenses lying along the $<111>$ directions; the dimension 
along the $\Gamma$-R line can be obtained from Fig. 3.  
The bottom panel shows an electron jack
at the $\Gamma$ point, an additional ellipsoidal surface (holes) at the
M point, and a squarish pillow (hole) at the X point.  

These Fermi surfaces can be compared with those presented by Chandra
{\it et al.}\cite{chandra}  Although there is much general similarity,
there are differences in detail that might have some importance.  
The wiffle ball and lenses show quite small differences (considering 
the complexity of this shape).  The electron jack has a decidedly different
shape but similar size, and one of our ellipsoid-like surfaces around
the M point becomes a much more convoluted surface in the result of
Chandra {\it et al.}  The associated differences in the two band structures,
which can be seen by comparing the band plots, reflect differences in
the accuracy of the respective code.  They have used a relatively new
code VENUS which makes it difficult to speculate on the origin of the
differences.  We have relaxed the positions of the Li and Pd atoms in
our calculation, however, and find that this change makes 
the extra $\Gamma$-centered
tiny jack in their calculation disappear in our results.

We have displayed the Fermi surfaces for the bands without spin-orbit
coupling.  As can be seen from the bottom panel of Fig. 3,  the 
actual Fermi surfaces will be only slightly more complicated due to
SOC (see Sec. \ref{Pdsoc}). 

\subsection{Spin-orbit Coupling without Inversion Symmetry}
\label{Pdsoc}
Interest in the effect of SOC in superconductors
without inversion symmetry has intensified very recently.  Li$_2$Pd$_3$B
with its T$_c$ = 8 K is the prime example at this time: Pd, and more especially Pt,
is heavy enough to make SOC an important consideration, and the
symmetry deviates seriously from inversion-symmetric.  As one
example of the magnitude of inversion-breaking: the B-Pd-B bond
angle, 180$^{\circ}$ in a perovskite, is only 162$^{\circ}$ in 
this system, and Pd (Pt) is where any appreciable SOC will arise.

The effect of SOC on the band structure of Li$_2$Pd$_3$B is
displayed in the lower panel of Fig. \ref{band}.  A surprise is 
evident: whereas the normal effect of SOC is to lift degeneracies at symmetry 
points/lines and introduce anticrossings among bands, in this case
the number of bands has doubled in addition to the usual effects.  This
doubling indicates that both 
spin-up and spin-down bands become mixed and appear on the same
band structure, a consequence of the lack of a center of inversion.  
When inversion symmetry is
present, the up- and down-spin bands are still mixed by SOC, 
but a degeneracy $k$ point by $k$ point is retained, 
the bands being spin conjugates of each other.
It is this degeneracy that the
lack of an inversion center removes, and it is the consequences of this
lack of symmetry that has attracted attention 
recently.\cite{samokhin,sergienko,samokhin2,frigeri0}
There is still a degeneracy $\varepsilon_{-k}
= \varepsilon_k$ due to time reversal; the first will have mostly 
spin-up character with some spin-down (say) and the other will 
the identical amount of spin-down with some spin-up.  Lack of inversion
does not require exotic pairing, since the states are still
time-reversed conjugates and can form singlet pairs (Anderson's theorem).  Triplet
pairing (or its generalization), on the other hand, 
is strongly suppressed, being driven to FFLO 
(Fulde-Farrel-Larkin-Ovchinnikov) type $q\neq 0$ pairing. 

Some quantitative estimate of an FFLO modulation in a `triplet'
state can be obtained.
Splittings near E$_F$ resulting from SOC + no inversion  range from zero
to 30 meV determined primarily by the amount of Pd
character.  Given the typical Fermi
velocity of 0.75 eV/($\pi/a$)$\approx$ 3$\times 10^7$ $cm/s$, the splitting of
Fermi surfaces can be up to $q =\Delta k \sim 0.04\pi/a$, with corresponding
modulation wavelength of $\sim$ 50$a$ $\sim$ 350~\AA.  The cost in
kinetic energy is $q^2/2m \sim$ 1 meV, which comparing to the superconducting 
condensation energy $\sim \Delta^2$/E$_F$ seems to kill any possibility of an
FFLO triplet state.  Moreover,
the observation of a Hebel-Slichter peak in the $^{11}$B spin-lattice
relaxation time\cite{nishiyama} strongly supports nodeless pairing.

\subsection{A$_g$ Phonon Modes}

\begin{figure}[tbp]
{\resizebox{8cm}{5cm}{\includegraphics{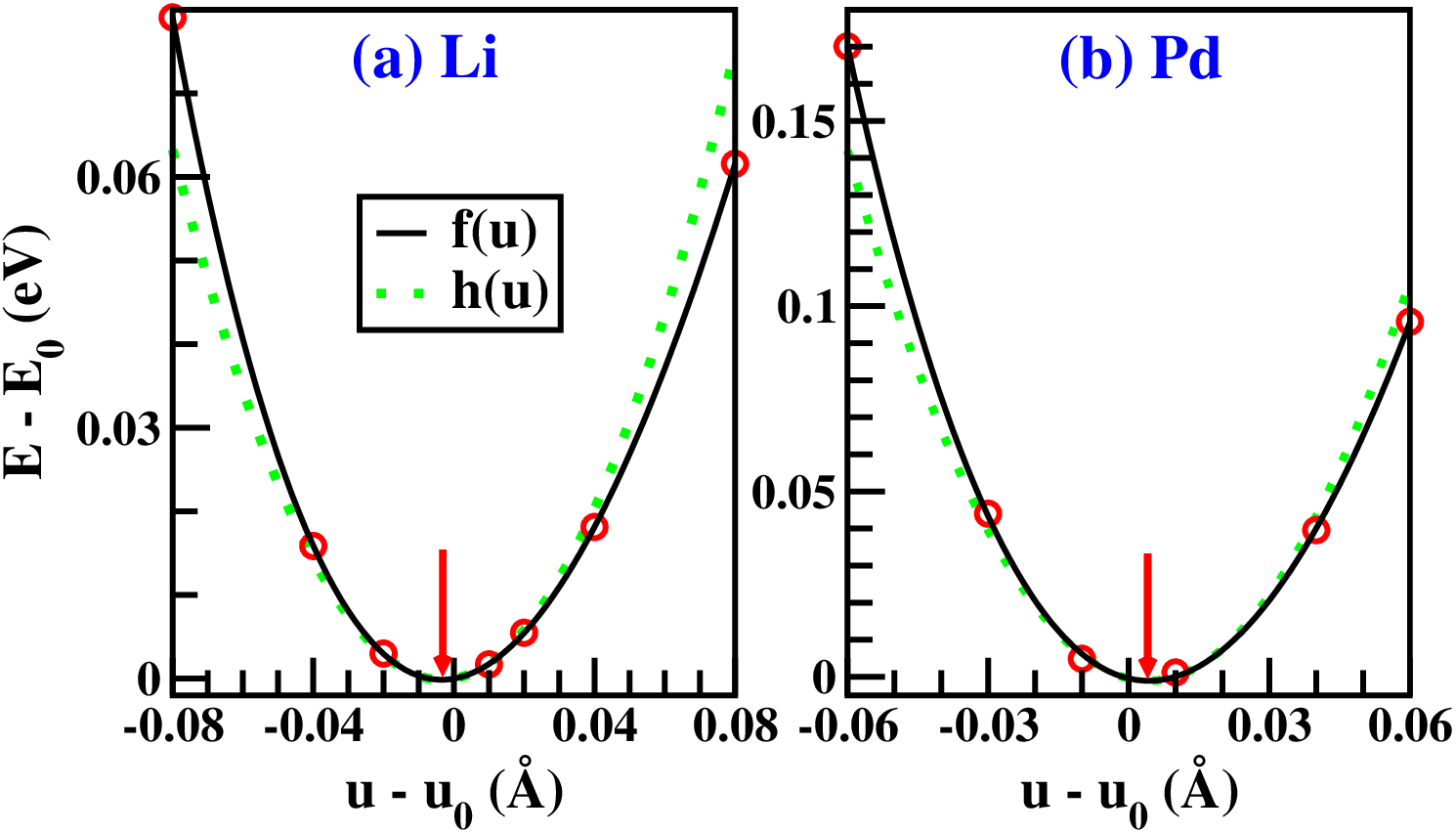}}}
\caption{(Color online)Energy change when 
(a) Li is displaced along the $<111>$ direction and (b) Pd along the
 $<011>$ direction.
 $u_0$ is the position displaced by $-0.06$ \AA~ for Li and $+0.01$ \AA~
 for Pd from the experimentally reported position.
 E$_0$ is energy for $u_0$.
 The fitting function f($u$) is given by $\varepsilon_0
 +a_2(u-u_o^\prime)^2+a_3(u-u_o^\prime)^3+a_4(u-u_o^\prime)^4$,
 and h($u$)=$\varepsilon_0+a_2(u-u_o^\prime)^2$ is a harmonic 
 approximated function of f($u$). $u_o^\prime$ is the equilibrium
 position denoted by the vertical arrows. }
\label{frozen}
\end{figure}

\begin{table}[bt]
\caption{Fitting parameters of the energy change for Li and Pd displacements.
 The parameters, a$_n$ for Li and b$_n$ for Pd (in eV/\AA$^n$), 
 are coefficients of the nth order displacement terms and c 
 (in eV/\AA$^2$) is the first order coupling term. The frequency $\omega$ 
 obtained from the harmonic approximation is given in meV.
 Li/Pd denotes a case where both atoms were displaced.
 The force constant $m\omega^2$, where $m$ is an atomic mass, is
 2.7 for Li and 5.8 for Pd (in eV/\AA$^2$).}
\begin{center}
\begin{tabular}{cccccc}\hline\hline
 displaced atom~ & ~a$_2$,~b$_2$~ &~ a$_3$,~b$_3$ ~ &~ a$_4$,~b$_4$~ 
                 &~ c ~&~ $\omega$~ \\ \hline
  Li  $<111>$   &10.8  & $-29.6$ & 73.5 &    & 40.4      \\
  Pd  $<011>$   &34.7  & $-83.6$ & 357.5&    & 15.0      \\ \hline
  Li/Pd         &10.6  & $-32.1$ & 151.8&    & 40.6      \\
                &33.2  & $-94.4$ & 583.0&16.9& 13.0      \\ \hline\hline
\end{tabular}
\end{center}
\label{table2}
\end{table}

Keeping the space group fixed, we can displace Pd along one $<011>$
direction (its variable internal coordinate) or (and) 
Li along a $<111>$ direction.
Figure \ref{frozen} shows the energy change for displacement of Li and Pd 
separately.
The data can be fit very well by expanding up to 4th order of
the displacement, and
the region near the equilibrium position is fit quite well
with the harmonic approximation.
The fitted constants are given in Table \ref{table2}. 
By displacing Li and Pd simultaneously, 
the two A$_g$ phonon frequencies and eigenvectors 
can be determined.
Considering only the harmonic term, the two frequencies are 
$\omega_1$=40.6 meV (96\% from Li and 4\% from Pd) and
$\omega_2$=13.0 meV (almost 50\% from the each type of atom).
Clearly
$\omega_2$ shows much more coupling effect, 13\% softening, whereas
$\omega_1$ has nearly negligible coupling,
consistent with the character of the phonon eigenvectors.
The value of $\omega_1$ is almost exactly equal to the
experimentally reported maximum value of Li metal,\cite{Bortolani}
whereas $\omega_2$
is much softer than the 30 meV maximum for Pd metal.\cite{Pd}

\subsection{Deformation Potential}
\begin{figure}[tbp]
\rotatebox{-90}{\resizebox{5cm}{8cm}{\includegraphics{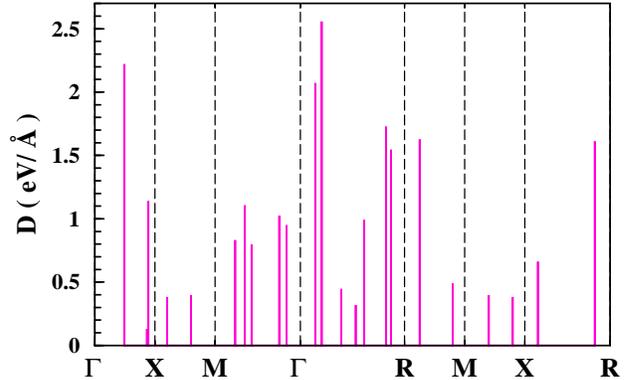}}}
\caption{(Color online)Distribution of the deformation potential ${\cal D}$ 
 at the Fermi level, when Pd moves along the $<011>$ direction. 
The vertical bar indicates the strength, and is placed at the 
positions where the bands cross the Fermi level.}
\label{D}
\end{figure}

A deformation potential ${\cal D}$ is defined as an energy shift with
respect to (periodic) atomic displacement.
Figure \ref{D} displays the strong $k$ variation of the deformation potential 
for selected bands at E$_F$ with respect to Pd motion 
along the $<011>$ direction. 
The value is large near the $\Gamma$ and R points but smaller near
the M and X points, reflecting the strong variation of the Pd character 
around the Fermi surface.  The variation of Pd/B character is pictured in the
fatband representations given in Fig. \ref{fat}, but the deformation
potential may not be simply related to the amount of Pd character. 
The averaged potential $\langle{\cal D}_{\tt Pd}\rangle$ for these points,
and its variance, is 
1.15($\pm 0.6$) eV/\AA, a value indicating
significant Pd contribution to electron-phonon coupling, as we show below.
The maximum value of about 4 eV/\AA =3.4 
$\langle{\cal D}_{\tt Pd}\rangle$ occurs for the flat band
lying just below E$_F$ at the $\Gamma$ point (not shown in 
Fig. \ref{D}), whereas the other
flat band at the X point has a value only comparable 
with $\langle{\cal D}_{\tt Pd}\rangle$.

For a physical feeling for the magnitude of the coupling, note that
the flat band at the $\Gamma$ point with large deformation potential
crosses E$_F$ 
when Pd is displaced by about $-0.03$ \AA~, while at the X point 
it requires a displacement of $-0.06$ \AA~ from the relaxed position. 
B $2p$ character at E$_F$
is consistent with the $^{11}B$ NMR data of Nishiyama and
coworkers.\cite{nishiyama}  
Unlike Pd, the Li deformation potential is small
(by an order of magnitude)
$\langle{\cal D}_{\tt Li}\rangle$= 0.11($\pm 0.06$) eV/\AA,  
with the maximum value of 0.27 eV/\AA~ occurring at the $\Gamma$ point.
The contribution of the deformation potentials to the electron-phonon
coupling constant can be measured by the quantity  
$\Lambda_i$=N(0)$\langle{\cal D}_i^2\rangle$/$\langle M_i\rangle$$\omega_i^2$, 
where $\langle M_i\rangle$ is an average mass of mode $i$
(for example, approximately $M_{\tt{Li}}$ for $\omega_1$
and a half of $M_{\tt{Pd}}$+$M_{\tt {Li}}$ for $\omega_2$).
Using the mean of the two values of $\langle{\cal D}_2\rangle$ 
for $\Lambda_2$, the ratio of $\Lambda_2/\Lambda_1$
is $30$, clearly indicating that the electron-phonon coupling 
is mainly from Pd contribution.  We have however not assessed possible
coupling from the B atom.  The large C isotope coefficient\cite{klimczuk}
$\alpha_C$ = 0.54 
in MgCNi$_3$ suggests that coupling to B indeed may be important.

We have used the Allen-Dynes equation\cite{allendynes} to assess 
these quantities in relation to the measured T$_c$ = 8 K.  Clear conclusions
are not possible because for T$_c$ of 8 K or less, the uncertain value of
the effective Coulomb repulsion $\mu^*$ introduces uncertainty.
We concentrate on the value $\lambda \sim$ 0.7, which is the specific
heat mass enhancement, and this value is also the ``mode $\lambda$'' for the
lower frequency A$_g$ mode.  Using a representative phonon frequency 
$\Omega$ = 13 meV (the A$_g$ mode), then $\mu^*$ = 0.10 gives T$_c$
= 5.4 K, while $\mu^*$=0.15 gives T$_c$ = 3.7 K.  Increasing $\Omega$
to 23 meV, the values become T$_c$($\mu^*$=0.10) = 9.6 K,
T$_c$($\mu^*$=0.15) = 6.0 K.  These values indicate that a mean
phonon energy of 20 meV with $\lambda$ = 0.7 may be necessary to
account for T$_c^{exp}$ = 8 K.

\section{The  P\lowercase{t} Analog}
The decrease of T$_c$ from 8 K to less than 3 K in the Pt analog provides a potentially
important clue into the mechanism of superconductivity, so we have looked at 
the differences in the electronic structure.  For Li$_2$Pt$_3$B, we used the 
experimentally reported structure
without relaxation,\cite{jung} since relaxation produced relatively small
difference in Li$_2$Pd$_3$B.
The valence orbitals of Pt were taken to be $4f5s5p6s6p5d$, with Li and B
being treated as before.

\begin{figure}[tbp]
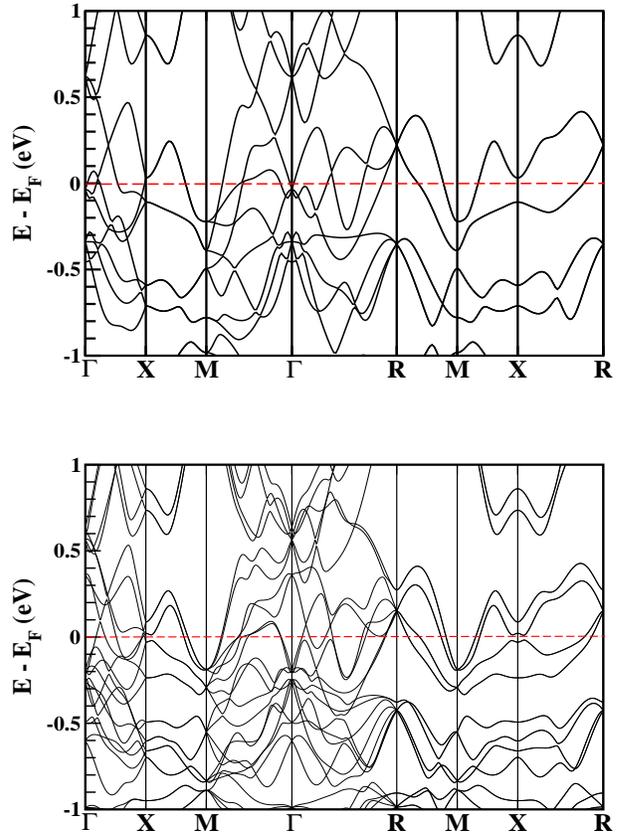

{\resizebox{8cm}{5cm}{\includegraphics{Fig8a.eps}}}
\vskip 10mm
{\resizebox{8cm}{5cm}{\includegraphics{Fig8b.eps}}}
\caption{Blowup of the Li$_2$Pt$_3$B band structure within 
1 eV of the Fermi energy, plotted identically to those for Li$_2$Pd$_3$B
in Fig. \ref{band}.  Top panel, without spin-orbit coupling; bottom panel,
spin-orbit included.  Even without spin-orbit, it is rather difficult 
to see any close correspondence
between the Li$_2$Pt$_3$B and Li$_2$Pd$_3$B (Fig. \ref{band}) bands; 
see text for further discussion of
differences.  The bottom panel reveals the strong effects in Pt arising from
coupling of the spins to real space motion of electrons.}
\label{band2}
\end{figure}

The resulting band structure is plotted in Fig. \ref{band2} on the same
scale as those of Li$_2$Pd$_3$B in Fig. \ref{band}, so direct comparison can
be made.  Given the same cell volume and the similarity of Pd and Pt in many
respects, it is somewhat surprising that the bands show so much difference. 
The Fermi level DOS actually increases, to 2.9 states/eV per formula unit.
 The differences, such as more bands in the region within 1 eV below
E$_F$, can be traced to the wider bandwidth, which also places more $d$
character at E$_F$.  The occupied bandwidth is 7.6 eV compared to 6.7 eV
in Li$_2$Pd$_2$B (Fig. \ref{dos}) and the $d$ bandwidth (region of 
large DOS) is $\sim$15\% wider and extends more strongly to E$_F$.

The bottom panel of Fig. 8 shows the consequence of SOC in the Pt compound.
Splittings near E$_F$ as large as 200 meV occur, for example, the band
below E$_F$ at the X point.  More generally, the splittings are perhaps 
on average a factor of two or so larger than for the Pd compound.  It can
be seen that fourfold degeneracies at the R point survive SOC, as do the
degeneracies at the M point, while the
threefold states at the $\Gamma$ point are split to doublet plus singlet.

The FPLO method, being atomic orbital based, can be used to provide a
Mulliken decomposition of charges.  While the atomic charges depend somewhat
on the choice of orbitals, it is possible that the differences in charges
for these two compounds can provide some insight.  The Mulliken effective
charges are, in the Li$_2$Pd$_3$B compound: Li, $-0.08$; Pd, $+0.17$; 
B, $-0.34$; and for the Li$_2$Pt$_3$B compound: 
Li, $+0.11$; Pt, $+0.02$; B, $-0.28$.  The main
difference is that Pt is effectively neutral while Pd is noticeably cationic.
These charges do not support the suggestions, discussed earlier,
that Li is cationic in Li$_2$Pd$_3$B, although it may  become
somewhat anionic in Li$_2$Pt$_3$B.

\section{Discussion and Summary}
Considering that the local coordination and the similar valences of
the constituents (Li$_2$ $\leftrightarrow$ Mg, Pd $\leftrightarrow$
Ni, B with one less electron than C) and that both compounds superconduct at
the same temperature of 8 K, it is worthwhile to consider
whether there is any realistic comparison with MgCNi$_3$.  In both
systems there is a nominal $d^{10}$ configuration of the transition
metal atom, but strong $d$ character remains at the Fermi surface.  The
volume per formula unit is 30\% larger in Li$_2$Pd$_3$B, much more
than expected from simple Ni$\rightarrow$Pd replacement and reflecting
the more open structure of Li$_2$Pd$_3$B.  Moreover, the band structures 
do not show much  resemblance (the four times larger cell
makes comparison difficult, however).  The DOS of MgCNi$_3$ is
dominated by a very high and very narrow peak in the density of
states 45 meV below E$_F$, derived from a very flat band all
around the M point of the Brillouin zone.  In Li$_2$Pd$_3$B there
is no analogous feature
(see Fig. \ref{band}).  There is a fairly flat band along the M-X line
that is cut, and hybridized with, a steeper band, as can be seen
in Fig. \ref{fat}, but there is no corresponding DOS peak.  
The value of N(0) is only 45\% of that of MgCNi$_3$ 
(per formula unit).\cite{johannes}
The band filling is, of course, one electron higher
in Li$_2$Pd$_3$B.  

Generally, we identify no close relationship
between the electronic structures of these two systems, and the 
differences are substantial.  The lack of inversion symmetry and
substantial (large) spin-orbit coupling in the Pd (respectively, Pt)
compound make the electronic structure much richer, with complicated
Fermi surfaces.  Comparison of the calculated
N(0) with the linear specific heat coefficient leads to a dynamical
mass enhancement $\lambda \sim 0.75$, which if due to electron-phonon
coupling is in the right range to account for T$_c$.  One particular 
Pd vibration is calculated to have a mode $\lambda$ of this same size.
However, the Li A$_g$ mode is found to be very weakly coupled.  This 
means that the electron-phonon coupling varies strongly throughout the
phonon spectrum, and most likely also across the Fermi surface.

The differences between Li$_2$Pd$_3$B and Li$_2$Pt$_3$B are strong 
enough that the origin of the difference in their values of T$_c$
is unclear.  Strong spin-orbit coupling together with the lack of
inversion symmetry, coupled with the observation of a Hebel-Slichter
coherence peak in the spin-lattice relaxation rate, strongly points to
conventional rather than exotic pairing, 
and we have begun to probe electron-phonon
coupling in this system.  MgCNi$_3$ has been found to have very
strong coupling to certain modes,\cite{savrasov} if the analogy 
to this compound is relevant.  If the coupling is primarily to the Pd (Pt),
the difference in masses leads to a decrease by $\sqrt{106/195}$ =
0.74, roughly half of the reduction factor that is needed.  A difference
in $\lambda$ of only 15\% would be required to give the further
reduction of T$_c$, and the electronic structure is different enough
to allow this possibility.

\section{Acknowledgments}
We thank K. Togano and H. Takeya for communicating experimental results 
on the linear specific heat coefficient
prior to publication.  Discussions with A. B. Shick about implications of lack of 
inversion symmetry have been very helpful.  This work was supported
by National Science Foundation grant No. DMR-0421810.


\begin{thebibliography}{10}

\bibitem{bauer} E. Bauer, G. Hilscher, H. Michor, Ch. Paul, E. W. Scheidt,
  A. Gribanov, Yu. Seropegin, H. No\"el, M. Sigrist, and P. Rogl, 
   Phys. Rev. Lett. {\bf 92},
   027003 (2004).

\bibitem{samokhin}K. V. Samokhin, E. S. Zijlstra, and S. K. Bose,
  Phys. Rev. B {\bf 69}, 094514 (2004).

\bibitem{sergienko}I. A. Sergienko and S. H. Curnoe, Phys. Rev. B {\bf 70},
  214510 (2004).

\bibitem{samokhin2}K. V. Samokhin, Phys. Rev. Lett. {\bf 94}, 027004 (2005).

\bibitem{frigeri0}P. A. Frigeri, D. F. Agterberg, and M. Sigrist,
  New J. Phys. {\bf 6}, 115 (2004).

\bibitem{pfleiderer} C. Pfleiderer, G. J. McMullan, S. R. Julian, and
  G. G. Lonzarich, Phys. Rev. B {\bf 55}, 8330 (1997).

\bibitem{frigeri}P. A. Frigeri, D. F. Agterberg, A. Koga, and M. Sigrist,
 Phys. Rev. Lett. {\bf 92}, 097001 (2004).

\bibitem{jeong}T. Jeong and W. E. Pickett, Phys. Rev. B {\bf 70}, 075114 (2004).

\bibitem{akazawa} T. Akazawa, H. Hidaka, T. Fujiwara, T. C. Kobayashi,
  E. Yamamoto, Y. Haga, R. Settai, and Y. Onuki, 
  J. Phys.: Condens. Matter {\bf 16}, L29 (2004).

\bibitem{togano1} K. Togano, P. Badica, Y. Nakamori, S. Orimo, 
  H. Takeya, and K. Hirata, Phys. Rev. Lett. {\bf 93}, 247004 (2004).

\bibitem{togano2} P. Badica, T. Kondo, T. Kudo, Y. Nakamori, S. Orimo,
  and K. Togano, Appl. Phys. Lett. {\bf 85}, 4433 (2004).

\bibitem{sardar} M. Saradar and D. Sa, Physica C {\bf 411}, 120 (2004).
Results of this paper appear to arise from an incorrect understanding of
the crystal structure.

\bibitem{jung} U. Eibenstein and W. Jung, J. Solid State Chem. {\bf 133},
  21 (1997).

\bibitem{togano3} T. Yokoya, T. Muro, I. Hase, H. Takeya, K. Hirata, 
  and K. Togano, Phys. Rev. B {\bf 71}, 092507 (2005).

\bibitem{nishiyama} M. Nishiyama, Y. Inada, and G.-Q. Zheng,
  Phys. Rev. B {\bf 71}, 220505(R) (2005).

\bibitem{chandra} S. Chandra, S. Mathi Jaya, and M. C. Valsakumar,
  cond-mat/0502525.

\bibitem{togano4} P. Badica, T. Kondo, and K. Togano, 
  J. Phys. Soc. Jpn. {\bf 74}, 1014 (2005).

\bibitem{he} T. He, Q. Huang, A. P. Ramirez, Y. Wang, K. A. Regan, 
   N. Rogado, M. A. Hayward, M. K. Haas, J. S. Slusky, K. Inumara, 
   H. W. Zandbergen, N. P. Ong, and R. J. Cava, Nature {\bf 411}, 54 (2001).

\bibitem{cava}R. E. Schaak, M. Avdeev, W.-L. Lee, G. Lawes, H. W. Zandbergen,
  J. D. Jorgensen, N. P. Ong, A. P. Ramirez, and R. J. Cava, J. Solid State
  Chem. {\bf 177}, 1244 (2004).

\bibitem{fplo1} K. Koepernik and H. Eschrig, Phys. Rev. B {\bf 59}, 1743 (1999).

\bibitem{fplo2} H. Eschrig, M. Richter, and I. Opahle, in
 {\it Relativistic Electronic Structure Theory - Part II: Applications},
  edited by P. Schwerdtfeger (Elsevier, Amsterdam, 2004), pp. 723-776.

\bibitem{Nf} Relaxation of the atomic position decreased 
  the value of N(0) by 4\%.
 Using the experimentally reported atomic position\cite{jung} and 
 140 irreducible $k$ points as was done in Ref. \cite{togano3}, we 
obtained the same value they reported. 

\bibitem{togano5} H. Takeya, K. Yamada, K. Yamaura, T. Mochiku, H. Fujii, T.
   Furubayashi, K. Hirata and K. Togano (private communication); 
   Physica C (in press).

\bibitem{Bortolani}V. Bortolani and G. Pizzichini, Phys. Rev. Lett.
 {\bf 22}, 840 (1969), and references therein. 

\bibitem{Pd}A. P. Miller and B. N. Brockhouse, Phys. Rev. Lett. {\bf 20},
  798 (1968).

\bibitem{klimczuk}T. Klimczuk and R. J. Cava, Phys. Rev. B {\bf 70},
  212514 (2004).

\bibitem{allendynes}P. B. Allen and R. C. Dynes, Phys. Rev. B {\bf 12}, 905 (1975).

\bibitem{johannes}M. D. Johannes and W. E. Pickett, Phys. Rev. B
 {\bf 70}, 060507(R) (2004).

\bibitem{savrasov}A. Yu. Ignatov, S. Y. Savrasov, and T. A. Tyson,
  Phys. Rev. B {\bf 68}, 220504(R) (2003).

\end{thebibliography}
\end{document}